\documentclass{PoS}

\title{The chemical evolution of the Milky Way} 
\ShortTitle{Milky Way} 
\author{\speaker{Francesca Matteucci}\\
        Dipartimento di Fisica,  Universit\`a di Trieste\\
        E-mail: \email{matteucci@oats.inaf.it}}
\author{Emanuele Spitoni\\
         Dipartimento di Fisica,  Universit\`a di Trieste\\
        E-mail: \email{spitoni@oats.inaf.it}}
\author{Donatella Romano\\
         I.N.A.F. Osservatorio Astronomico di Bologna\\
        E-mail: \email{donatella.romano@oabo.inaf.it}}
\author{Alvaro Rojas-Arriagada\\
Instituto de Astrof\'{i}sica, Facultad de F\'{i}sica, Pontificia
Universidad Cat\'olica de Chile, Av. Vicu\~na Mackenna 4860, Santiago, Chile

Millennium Institute of Astrophysics, Av. Vicu\~{n}a Mackenna 4860,
782-0436 Macul, Santiago, Chile \\
E-mail: \email{arojas@astro.puc.cl}}

\abstract{
 We will discuss some highlights concerning the chemical evolution of our Galaxy, the Milky Way. First we will describe the main ingredients necessary to build a model for the chemical evolution of the Milky Way. Then we will illustrate some Milky Way models which includes detailed stellar nucleosynthesis and compute the evolution of a large number of chemical elements, including C, N, O, $\alpha$-elements, Fe and heavier. The main observables and in particular the chemical abundances in stars and gas will be considered. A comparison theory-observations will follow and finally some conclusions from this astroarchaeological approach will be derived. }

\FullConference{Frontier Research in Astrophysics - II\\
         23-28 May 2016\\
         Mondello (Palermo), Italy}

\begin{document}

\section{Introduction}

In the last years a great deal of spectroscopic data concerning a very large number of Galactic stars has appeared in the literature. Many surveys have released chemical abundances for hundreds thousands of stars in the Galaxy: for example RAVE (Steinmetz et al. 2006), SEGUE-1 (Yanny et al. 2009), SEGUE-2 (Rockosi et al. 2009), ARGOS (Freeman et al. 2013), LAMOST (Cui et al. 2012), Gaia-ESO (Gilmore et al. 2012), GALAH (Zucker et al. 2012), APOGEE (Majewski et al. 2015) and the AMBRE project (de Laverny et al. 2013).
It is therefore very important to construct theoretical models able to reproduce the observed abundance patterns. Galactic chemical evolution models had a great start thanks to the pioneering work of Beatrice Tinsley (1941-1981) who established the basis of this important field. Galactic chemical evolution describes how the gas and its chemical composition evolve in galaxies of different morphological type.  As it is well known, during the Big Bang only light elements were formed (H, He, D, Li), while all the other species from carbon (elements with $A> 12$) were built inside the stars: stars produce chemical elements and restore them into the ISM, out of which new stars will born. This is the chemical evolution process. Many chemical evolution models have been proposed in the last 40 years and have demonstrated how important is to relax the hypothesis of instantaneous recycling approximation to compute the evolution of the abundances of different chemical elements, as well as how important are the relative contributions of Type Ia and core-collapse supernovae (SNe) to the chemical enrichment. 
In particular, it has been established the principle of the ``time-delay model'', which allows us to interpret the abundance patterns measured in stars: for example, the [$\alpha$/Fe] versus [Fe/H] relation and how it is expected to vary in different galaxies (see Matteucci 2012 for a review on the subject). The time-delay model interprets the observed trends as due to the different timescales on which different chemical elements are produced, as for example $\alpha$-elements and iron. The $\alpha$-elements are produced on short timescales by core-collapse SNe, whereas Fe is produced on longer timescales (with a delay) by Type Ia SNe. By means of such an interpretation we are able to establish the timescales of the formation of galaxies and of separate galactic components, such as halo, disk and bulge in the Milky Way. The stars in each component show different abundance patterns, indicating different histories of star formation. In fact, through the chemical abundances we can infer the formation and evolutionary history of galaxies and their components, and this approach is known as ``astroarchaeological approach''.
In this paper, we will concentrate on the evolution of the Milky Way and on what we have learned up to now. We will present model results compared to the most important and recent observations. From these comparisons we will derive the timescales for the formation of the different Galactic components as well as constraints on stellar nucleosynthesis.

\section {The chemical evolution model for the Milky Way} 

The main ingredients to build a chemical evolution model are: i) initial conditions, ii) the history of star formation, namely the star formation rate (SFR) and the initial mass function (IMF), iii) the stellar yields, iv) gas flows in and out the galaxy.
The most common parametrization of the SFR is that of Kennicutt (1998), that we also adopt:
\begin{equation}
\psi(t)=\nu \sigma_{gas}^{1.4},
\end{equation}
where $\nu$ is the efficiency of star formation and it should be tuned to reproduce the present time SFR in
the object one is modeling.
The IMF is normally a power law. The most commonly adopted IMFs are the one of Salpeter (1955), Scalo (1986), Kroupa et al. (1993), Kroupa (2001), Chabrier (2003).
The stellar yields are very important ingredients for chemical evolution, the nucleosynthesis prescriptions we will adopt in our model are described below.

\subsection{Nucleosynthesis prescriptions}

For chemical evolution models, the nucleosynthesis prescriptions and the implementation of the yields in the model are fundamental ingredients. In this work, we adopt the same nucleosynthesis prescriptions of model 15 of Romano et al. (2010), where a detailed description of the adopted yields can be found.
\\As regard to the computation of the stellar yields, one has to distinguish between different mass ranges, as well as single stars versus binary systems:
\begin{itemize}
\item low- and intermediate-mass stars (0.8 $M_{\bigodot}$-8 $M_{\bigodot}$), which are divided in single stars and binary systems. Binary systems formed by a white dwarf and a low or intermediate mass star companion can originate either Type Ia SNe or novae (when the companion is a low mass star),
\item massive stars (M $>$ 8 $M_{\bigodot}$).
\end{itemize}

\subsubsection{Low- and intermediate-mass stars}

\begin{itemize}
\item \textit{Single stars.} The single stars in this mass range contribute to the galactic chemical enrichment through planetary nebula ejection and quiescent mass loss along the giant and asymptotic giant branches. They enrich the ISM mainly in He, C, and N and heavy s-process elements. They can also produce non-negligible amounts of $^7$Li. For these stars, which end their lives as white dwarfs, we adopt the prescriptions of Karakas (2010).
\item \textit{Type Ia SNe.} Type Ia SNe are thought to originate from carbon deflagration in C-O white dwarfs in binary systems. Type Ia SNe contribute a substantial amount of iron (0.6 $M_{\bigodot}$ per event) and non negligible quantities of Si, S and Ca. They also contribute to other elements, such as O, C, Ne, Mg, but in negligible amounts compared to the masses of such elements ejected by massive stars. The adopted nucleosynthesis prescriptions are from Iwamoto et al. (1999).
\end{itemize}

\subsubsection{Massive stars}

Massive stars are the progenitors of Type II, Ib and Ic SNe and they are known as ``core-collapse SNe''. In particular, the SNe Ib,c are the explosion of stars with masses larger than $\sim 30 M_{\odot}$,  whereas SNe II originate from stars in the mass range $8< M/M_{\odot}<30$. If the explosion energies are significantly higher than 10$^{51}$ erg, hypernova events may occur (SNe Ic). For core-collapse SNe, we adopt up-to-date stellar evolution calculations by Kobayashi et al. (2006) for the following elements: Na, Mg, Al, Si, S, Ca, Sc, Ti, Cr, Mn, Co, Ni, Fe, Cu and Zn. As for the He and CNO elements, we take into account the results of Geneva models for rotating massive stars (see Romano et al. (2010) for references).

\subsection{The model for halo and disks}
The results we will present relative to the Galactic halo, thick and thin disk, are based on the two-infall model of Chiappini et al. (1997). In this model we assume that the halo and thick disk formed out of a relatively fast episode of gas accretion occurring on a timescale no longer than 2 Gyr, whereas the thin disk formed out of another accretion episode occurring on much longer timescales: an inside-out formation with the solar region forming on a timescale of 7 Gyr.
The assumed progenitors for Type Ia SNe are white dwarfs in binary systems, in particular the single-degenerate model (see Matteucci \& Greggio 1986; Matteucci \& Recchi, 2001).

The IMF adopted for halo and disks is the Scalo (1986) one. A gas threshold of $7 M_{\odot}pc^{-2}$ for star formation is assumed. This threshold produces naturally a gap in the star formation between the end of the halo-thick disk phase and the thin disk phase.

\begin{figure}[h]
\centering \includegraphics[scale=0.6]{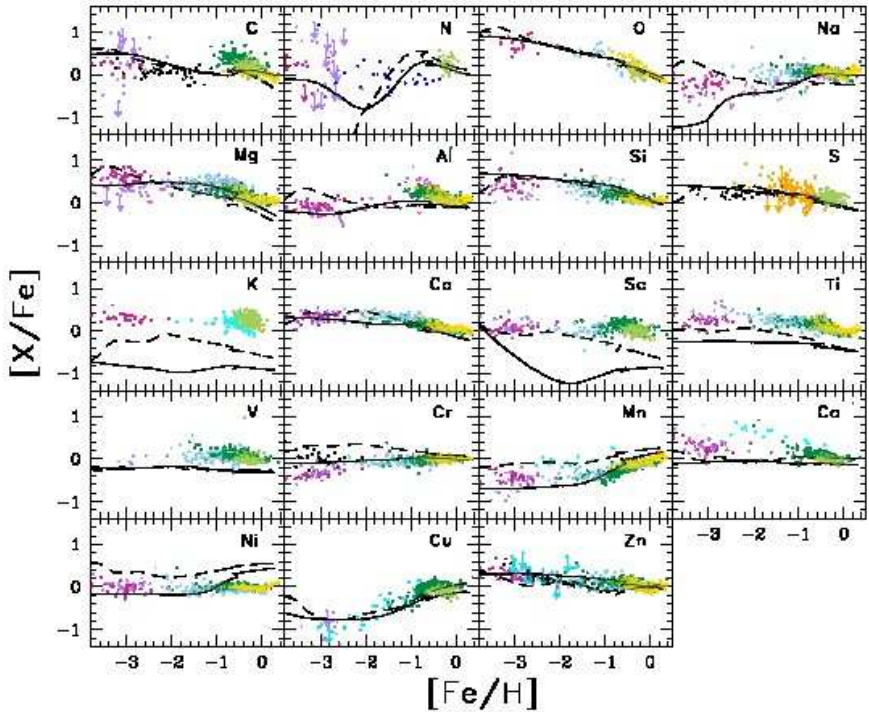} 
\caption{Predicted and observed [X/Fe] vs. [Fe/H] relations for the solar vicinity. Results and figure from Romano et al. (2010).The model results refer to two different sets of yields. The best set is
that represented by the solid lines with the nucleosynthesis
  prescriptions described here. The adopted IMF in both cases is that of Kroupa et al. (1993).
}
\label{abundances}
\end{figure}

In Fig.1 we show the results obtained for the yields discussed above (continuous line) as well as for older sets of yields (dashed line), as discussed in Romano et al. (2010). It is clear from Figure 1 that the trend of some chemical elements is well reproduced, whereas there are some elements whose yields should be strongly revised (i.e. K, Sc, Ti).

Another model, after that of Chiappini et al. (1997), has been suggested for the Milky Way (Micali et al 2013), where the thick disk phase was considered separately from the halo and thin disk formation. This model has been called ``three-infall model''. In this scenario, the halo formed very fast during a first accretion episode, then followed the accretion episode forming the thick disk, on a time scale short but longer than the halo ($\sim 2$ Gyr)

In Figure 2 we show the predicted [O/Fe] vs. [Fe/H] by the three-infall model,
compared to data relative to the halo, thick and thin disk. The halo, thick and thin disk star data are indicated in the Figure but a clear separation between thick and thin disk stars is not evident, so that strong conclusions cannot be derived, except that the thick disk stars, being $\alpha$-enhanced should have formed in a faster process than the thin disk stars. 

\begin{figure}[h]
\centering \includegraphics[scale=0.4]{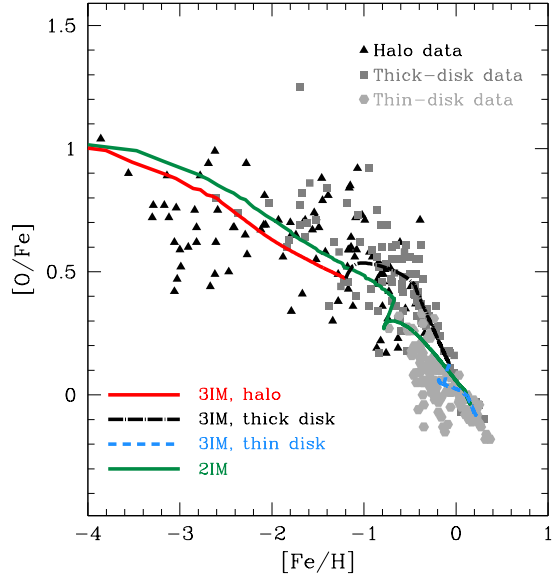} 
\caption{The [O/Fe] vs. [Fe/H] predicted by Micali et al (2013) 
  compared with data. The predictions for the halo are in red, those for the
  thick disk in black and in blue those for the thin disk. A comparison with
the two-infall model (green line) is also shown.
}
\label{micali}
\end{figure}

Recently, many data have appeared on the thick and thin disk stars (Hayden et
al. 2015; Rojas-Arriagada et al. 2017) where a clear separation is evident.
In Figure 3 we show some recent data from Gaia-ESO survey (Rojas-Arriagada
et al. 2017) where the thick and thin disk stars are
clearly separated in the plot [Mg/Fe] vs. [Fe/H]. In particular, it looks like
if the thick and thin disk formed on two different timescales
but in parallel and not sequentially, as in the model described above.

\begin{figure}[h]
  \centering \includegraphics[scale=0.8]{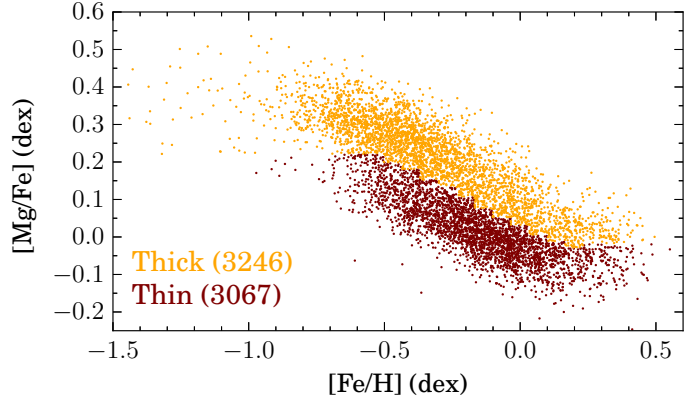}
  \caption{[Mg/Fe] vs. [Fe/H] in stars in the thick and thin disks from Gaia-ESO data from Rojas-Arriagada et al. (2017). The two parallel sequences for thin and thick disk are evident from the Figure.
}
\label{disks}
\end{figure}

From the models presented above and comparison with the data, we can conclude that the halo formed on a timescale of less than 1 Gyr, the thick disk on a timescale of 1-2 Gyr, no longer, as proven by the observed enhanced abundances of $\alpha$-elements relative to Fe, and the thin disk in the solar vicinity formed on a much longer timescale (7 Gyr).

\begin{figure}[h]
 \centering\includegraphics[scale=0.8]{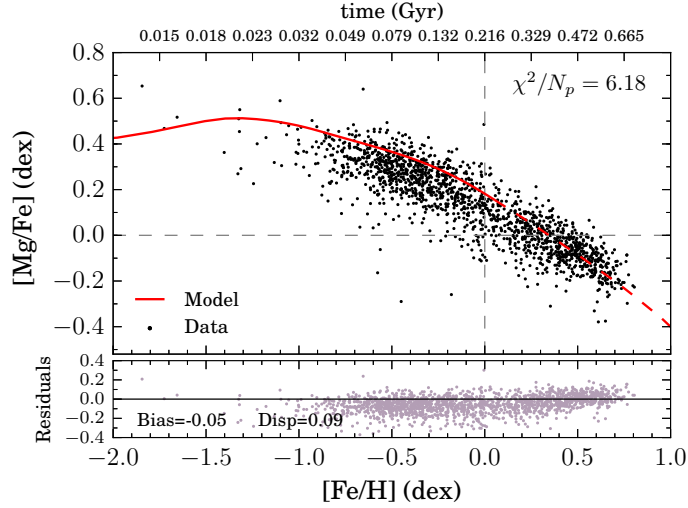}
\caption{[Mg/Fe] vs. [Fe/H] in Galactic bulge stars from Rojas-Arriagada et al. (2017) compared to model predictions. The model assumes a fast bulge formation and a Salpeter IMF, as described in the text.
}
\label{rojas}
\end{figure}

\subsection{The Galactic bulge}
The Galactic bulge shows a stellar  metallicity distribution function peaked at higher metallicity than the G-dwarfs in the solar vicinity. This means that the bulge stars formed faster than those in the solar vicinity.
Hill et al. (2011) suggested that the Galactic bulge contains two main stellar populations: a) a classical bulge stellar population called metal poor and b) a metal rich population which can be the result of star formation induced by the Galactic bar or be inner disk stars.
The most recent data on abundance ratios in bulge stars (coming from high-resolution, high-S/N spectroscopic data)
are from the Gaia-ESO survey. In Figure 4 we show these data together with the prediction of a bulge model relative to the classical bulge population, as described in Grieco et al. (2012), assuming a very fast SFR with star formation efficiency $\nu=25 Gyr^{-1}$ and timescale of gas accretion $\tau=0.1$ Gyr. As one can see, the model prediction fits well the data, thus confirming previous papers (Matteucci \& Brocato 1990; Ballero et al. 2007; Cescutti et Matteucci 2011; Grieco et al. 2012) suggesting a very fast bulge formation, at least for the main population (the endemic metal-poor classical one).
The stellar metallicity distribution function for bulge stars, where there is evidence of two populations (Hill et al 2011; Gonzalez et al 2015, Zoccali et al 2017, GIBSsurvey; Rojas-Arriagada 2014,2017, Gaia-ESO survey; Schultheis et al 2017,APOGEE survey, in contrast with Ness et al 2013, ARGOS survey) is shown in Figure 5, with the model results of Grieco et al. (2012) compared to the observations.
\begin{figure}[h]
\centering \includegraphics[scale=0.4]{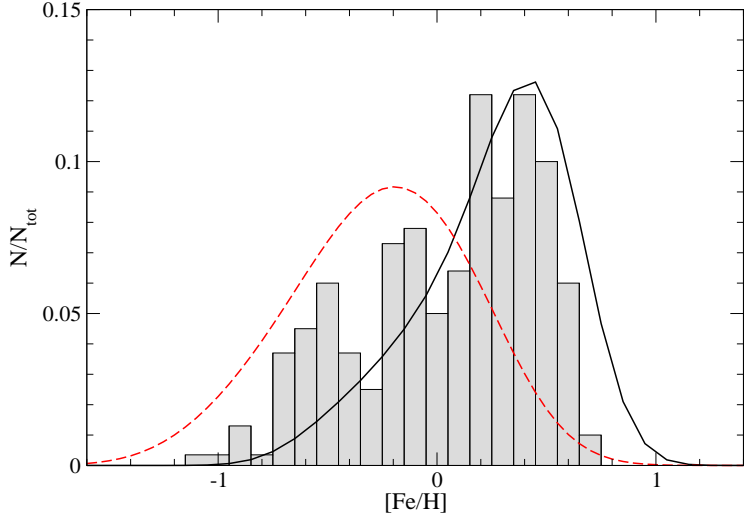}
\caption{Two stellar population in the Galactic bulge. Data (histogram) are from Hill et al. (2011); models (curves) are from Grieco et al. (2012). The red dashed curve refers to the so-called metal poor classical bulge population, while the black continuous curve refers to the metal rich population.
}
\label{grieco}
\end{figure}

\subsection{Abundance gradients along the Galactic disk}

Different models for the evolution of the entire thin disk  have appeared in the past years. These models include radial gas flows and stellar migration (Schoenrich \& Binney 2009; Spitoni \& Matteucci 2011; Spitoni et al. 2015; Kubryk et al. 2015). 
Radial gas flows are important for the creation of an abundance gradient in the gas along the thin disk. Other important parameters are: i) the inside-out formation of the disk, ii) the existence of a gas threshold for star formation.
In Figure 6 we show the gradients predicted by Spitoni \& Matteucci (2011).
It is clear from the comparison with O abundance measured in  HII regions, planetary nebulae and O, B stars that the model with inside-out formation and radial gas flows with variable speed best reproduces the observed gradient.

\begin{figure}[h]
\centering \includegraphics[scale=0.4]{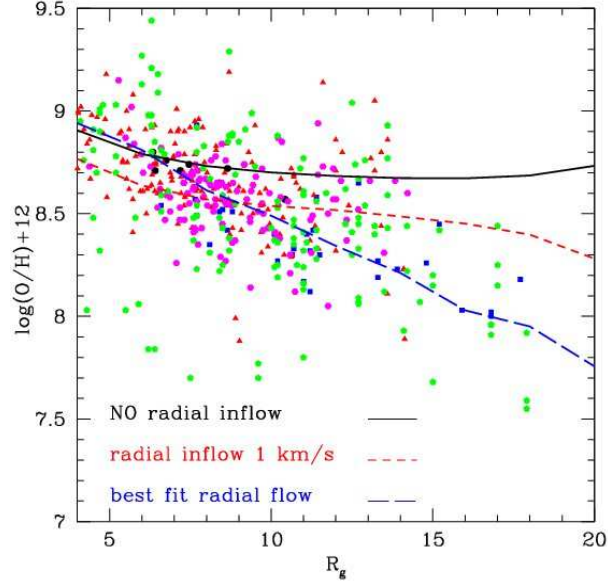}
 \caption{Comparison between data and models about abundance gradients along the Galactic thin disk. References to the data can be found in Spitoni \& Matteucci (2011). The black continuous line is  the result of a model without radial gas flows but including inside-out disk formation and gas threshold in the star formation. The red short-dashed line is the same model with radial flows at constant speed (1 km/sec) and the blue long-dashed line is the same model with radial flows with variable speed. This model is clearly the best one.
}
\label{gradients}
\end{figure}

Models with a constant timescale of gas accretion with galactocentric distance cannot predict any gradient. Models without inside-out but with radial gas flows
can in principle produce a gradient.
On the other hand, a model with inside-out formation without a gas threshold for star formation and radial gas flows can still produce a gradient in reasonable agreement with the data, but only for the very inner disk regions, as it is evident from Figure 6. For the outermost regions there is practically no gradient.

\section{Summary and Conclusions}

Galactic astroarchaeology is a useful tool to infer the timescales for the formation of the various Galactic components: halo, thick, thin disk and bulge. To do that one compares the predicted and observed abundance patterns. From the discussion above we can suggest the following:

\begin{enumerate}

\item The halo phase lasted no longer than 0.5-1.0 Gyr. This is dictated by the large overabundances of $\alpha$-elements observed in halo stars. The thick disk also formed quickly, although on a longer timescale than the halo. Also thick disk stars show overabundances of $\alpha$-elements and their metallicity distribution function can be well reproduced if a timescale no longer than 2 Gyr is assumed (Micali et al. 2013).

\item The thin disk instead formed on a much longer timescale and inside-out. In particular, the inner regions formed on timescales of 1-2 Gyr, whereas the solar region took at least 7 Gyr, as suggested by the metallicity distribution function of the G-dwarfs, and the outermost regions ($R>14$ kpc) over 10 Gyr.

\item Probably the star formation stopped briefly between the formation of the halo and thick disk and between the thick and thin disk.
Haywood et al. (2016), by considering APOGEE data (Hayden et al. 2015), concluded that there was a quenching in the star formation at the end of the thick disk phase. Kubryk et al. (2015) suggested instead that the thick disk is the result of stellar migration: they concluded that the thick disk is the early part of the Milky Way disk. Very recent data from Gaia-ESO survey (Rojas-Arriagada et al. 2017) seem to suggest that the thick and thin disk formed in parallel. In such a case we would not expect any halt in the star formation, but the timescales of formation of the two disks should be the same as suggested above.

\item The Galactic bulge formed very quickly, on a timescale of 0.3-0.5 Gyr, at least the bulk of its stars, the classical bulge stellar population. A
population of stars whose formation was triggered by the bar (those
participating in the X-shape bulge) is present, outnumbering at the
metal-rich part of the bulge MDF a possibly small fraction of endemic
metal-rich bulge stars belonging to the initial classical population. The
stellar metallicity distribution is showing in fact a bimodal distribution.

\item Finally, Galactic abundance gradients can arise as a result of the inside-out formation of the thin disk coupled with radial flows with a speed variable with galactocentric distance (Spitoni \& Matteucci, 2011).

\end{enumerate}

\end{document}